\documentstyle [12pt]{article}
\begin{document}
\baselineskip 20.0 pt
\par
\mbox{}
\vskip -1.25in\par
\mbox{}
 \begin{flushright}
\makebox[1.5in][1]{ITP-SB-92-66}\\
\makebox[1.5in][1]{UU-HEP-92/23}\\
\makebox[1.5in][1]{December 1992}
\end{flushright}
\vskip 0.25in

\begin{center}
{\large Many Boson Realizations of Universal Nonlinear $W_{\infty}$-Algebras,
Modified KP Hierarchies and Graded Lie Algebras}\\
\vspace{40 pt}
{Feng Yu}\\
\vspace{20 pt}
{{\it Institute for Theoretical Physics, State University of New York}}\\
{{\it Stony Brook, New York 11794, U.S.A.}}\\
{{\it and}}\\
{{\it Department of Physics, University of Utah}}\\
{{\it Salt Lake City, Utah 84112, U.S.A.}}\\
\vspace{40 pt}
{{\large Abstract}}\\
\end{center}
\vspace{10 pt}

An infinite number of free field realizations of the universal nonlinear
$\hat{W}_{\infty}^{(N)}$ ($\hat{W}_{1+\infty}^{(N)}$) algebras, which
are identical to the KP Hamiltonian structures, are obtained in terms of
$p$ plus $q$ scalars of different signatures with $p-q=N$. They are
generalizations of the Miura transformation, and naturally give rise to the
modified KP hierarchies via corresponding realizations of the latter.
Their characteristic Lie-algebraic origin is shown to be the graded $SL(p,q)$.

\newpage

Extended $W$-algebras arise as fundamental symmetries or constraints in 2d
conformal field theories and low-dimensional string theories. In these
discoveries, a crucial link, elucidating the rather mathematical
formalisms in terms of more intuitive physical constituents, has been
the $W$-algebras' free field realization and moreover their intrinsic
connection to simple Lie algebras. On the other hand, $W$-algebras are
isomorphic to the Hamiltonian structures of the integrable KP (KdV) systems,
hence their realization naturally provides a field-theoretical construction
of the relevant hierarchies. Two important examples are the Miura
transformation in the KdV context [1] for $W_{N}$ [2] and the two boson
$\hat{W}_{\infty}$ realization [3] in the KP basis. The former expresses
the $W_{N}$ generators, defined as the coefficient functions $q_{r}(z)$
of the differential KdV operator due to the identification of $W_{N}$
with the second KdV Hamiltonian structure, in terms of the chiral currents
of $N$ free bosons $\vec{j}(z)=\vec{\phi}'(z)=(j_{1},j_{2},\cdots ,j_{N})$:
\begin{eqnarray}
Q = D^{N} + \sum_{r=0}^{N-2}q_{r}D^{r} = \prod_{i=1}^{N}(D+\vec{h}_{i}
\cdot\vec{j}), ~~ D\equiv \partial/\partial z
\end{eqnarray}
where $\vec{h}_{i}$ are the fundamental weight vectors of $SL(N)$ satisfying
$\vec{h}_{i}\cdot\vec{h}_{j}=\delta_{ij}-\frac{1}{N}$. It leads to the
solvable $W$-minimal models [4]. Meanwhile, it allows a
$\vec{j}(z)$-constitution of the KdV hierarchies, which have been called the
modified KdV [5]. In the latter case, the generators $u_{r}(z)$ of
$\hat{W}_{\infty}$ [6], a general homogeneous nonlinear deformation of
$W_{\infty}$ [7] and isomorphic to the second Hamiltonian
structure [8] of the KP hierarchy [9], are realized by a pair of free bosonic
currents $\bar{j}(z)=\bar{\phi}'(z)$, $j(z)=\phi'(z)$ through the
pseudo-differential KP operator:
\begin{eqnarray}
L = D+\sum_{r=0}^{\infty}u_{r}D^{-r-1} = D+\bar{j}\frac{1}{D-(\bar{j}+j)}j
\end{eqnarray}
with
\begin{eqnarray}
D^{-1}f(z) = \sum_{l=0}^{\infty}(-1)^{l}(\partial^{l}_{z}f)D^{-l-1}.
\end{eqnarray}
It was based on eq.(2) that a $\hat{W}_{\infty}$ current algebra [10,3]
and further an infinite set of commuting $\hat{W}_{\infty}$ charges [11,12]
were uncovered in the $SL(2,R)/U(1)$ coset model. Besides, KP hides much
richer algebraic structure. By taking the powers of (2) or the extention of
(1) to negative powers
\begin{eqnarray}
L^{N} = D^{N} + \sum_{r=-N+1}^{\infty}v_{r}D^{-r-1}
\end{eqnarray}
as basic ingredients, nonequivalent KP Hamiltonian structures
$\hat{W}_{\infty}^{(N)}$ are generated by $v_{r}(z)$ with spin, $s=r+N+1$,
ranging from 2 to $\infty$, which may even be formally generalized to
a family of nonlinear $W$-infinity algebras [13]. However, there lack both
field-theoretical realization and Lie-algebraic interpretation of
$\hat{W}_{\infty}^{(N)}$.

In the following, we present an infinite number of free field realizations for
$\hat{W}_{\infty}^{(N)}$, including $\hat{W}_{\infty}\equiv
\hat{W}_{\infty}^{(1)}$, in terms of $p$ bosons of one signature plus $q$
bosons of the opposite signature with $p-q=N$. They are generalizations of the
Miura transformation (1) and the two boson $\hat{W}_{\infty}$ realization (2).
Moreover, under these realizations, $\hat{W}_{\infty}^{(N)}$, or rather
$\hat{W}_{1+\infty}^{(N)}$ by incorporating a spin-1 current, are reducible
with each other, thus are all universal nonlinear $W$-algebras.
Correspondingly, we obtain the many boson realizations of KP, defining the
modified KP hierarchies. We further show, via our bosonic realizations, that
the characteristic Lie algebra associated with $\hat{W}_{\infty}^{(N)}$ is
the graded $SL(p,q)$ [14]. These results have played a key role in justifying
the $\hat{W}_{\infty}^{(N)}$-constraints on the recently constructed $D>2$
topological strings [15].

Let us proceed as usual in the KP basis in which $\hat{W}_{\infty}^{(N)}$ can
be most conveniently exhibited. To be generic, we include the $v_{-N}D^{N-1}$
term in the $N$-th KP operator (4), which is equivalent to extending
$\hat{W}_{\infty}^{(N)}$ to $\hat{W}_{1+\infty}^{(N)}$. The complete
structure of $\hat{W}_{1+\infty}^{(N)}$
\begin{eqnarray}
{\{v_{r}(z), v_{s}(z')\}} = k_{rs}^{(N)}(z)\delta(z-z')
\end{eqnarray}
may be extracted from the bilinear Hamiltonian form
\begin{eqnarray}
K(P) \equiv (L^{N}P)_{+}L^{N}-L^{N}(PL^{N})_{+} = \sum_{r=-N}^{\infty}
k_{rs}^{(N)}p_{s}D^{-r-1}
\end{eqnarray}
where $P=\sum D^{s}p_{s}$ is an arbitrary pseudo-differential operator
and the subscript $+$ denotes the purely differential part, and the infinite
dimensional operator matrix $k_{rs}^{(N)}$ were given explicitly in
closed form in [12,13]. Consider now the free bosonic currents $j_{i}(z)$
$(i=1,\ldots,p)$ and $\tilde{j}_{l}(z)$ $(l=p+1,\ldots,p+q)$ satisfying the
Poisson brackets
\begin{eqnarray}
& & {\{j_{i}(z),j_{j}(z')\}} = \delta_{ij}\partial_{z}\delta(z-z'),
{}~~~ {\{\tilde{j}_{l}(z),\tilde{j}_{k}(z')\}} = -\delta_{lk}
\partial_{z}\delta(z-z'), \nonumber\\
& & {\{j_{i}(z),\tilde{j}_{l}(z')\}} = 0.
\end{eqnarray}
With $p-q=N$ but $q$ (or $p>N$) remaining arbitrary positive integers,
there exist an infinite number of transformations of the $N$-th KP operator:
\begin{eqnarray}
L^{N} = D^{N} + \sum_{r=-N}^{\infty}v_{r}D^{-r-1}
= \prod_{i=1}^{p}(D+j_{i})\prod_{l=p+1}^{p+q}(D-\tilde{j}_{l})^{-1}.
\end{eqnarray}
In this way, the $\hat{W}_{1+\infty}^{(N)}$ currents $v_{r}$ are functions
of $j_{i}$, $\tilde{j}_{l}$ and their derivatives, and the
$\hat{W}_{1+\infty}^{(N)}$ brackets (5) are realized according to (7).

To prove the theorem, let us evaluate the brackets between two arbitrary
functionals $\oint f(L^{N})$ and $\oint g(L^{N})$ with $L^{N}$ in the form
of (8),
\begin{eqnarray}
{\{\oint f, \oint g\}} = \oint (\sum_{i=1}^{p}\frac{\delta f}{\delta j_{i}}
(\frac{\delta g}{\delta j_{i}})'- \sum_{l=p+1}^{p+q}
\frac{\delta f}{\delta\tilde{j}_{l}}(\frac{\delta g}{\delta\tilde{j}_{l}})'),
\end{eqnarray}
where $\delta f/\delta j$ stands for usual variational derivative. To be
concise, introduce
\begin{eqnarray}
\frac{\delta f}{\delta L^{N}} = \sum_{r=-N}^{\infty}D^{r}
\frac{\delta f}{\delta v_{r}},
\end{eqnarray}
followed by
\begin{eqnarray}
\oint Res (\delta L^{N}\frac{\delta f}{\delta L^{N}}) = \oint\sum_{r}
\delta v_{r}\frac{\delta f}{\delta v_{r}} = \oint (\sum_{i}\delta j_{i}
\frac{\delta f}{\delta j_{i}} + \sum_{l}\delta \tilde{j}_{l}
\frac{\delta f}{\delta\tilde{j}_{l}})
\end{eqnarray}
where $Res$ means the coefficient of the $D^{-1}$ term, so that
\begin{eqnarray}
\frac{\delta f}{\delta j_{i}} &=& Res(\prod_{j=i+1}^{p}(D+j_{j})
\prod_{l=p+1}^{p+q}(D-\tilde{j}_{l})^{-1}\frac{\delta f}{\delta L^{N}}
\prod_{j=1}^{i-1}(D+j_{j})), \nonumber\\
\frac{\delta f}{\delta \tilde{j}_{l}} &=& -Res(\prod_{k=p+l}^{p+q}
(D-\tilde{j}_{k})^{-1}\frac{\delta f}{\delta L^{N}}
\prod_{i=1}^{p}(D+j_{i})\prod_{k=p+1}^{p+l}(D-\tilde{j}_{k})^{-1}).
\end{eqnarray}
Replacing $(\delta g/\delta j_{i})'$ by the commutator ${[D+j_{i},
\delta g/\delta j_{i}]}$, etc., and substituting (12) into (9), we have
\begin{eqnarray}
{\{\oint f, \oint g\}}
&=& \oint \sum_{j=1}^{p}Res\{\prod_{i=j+1}^{p}(D+j_{i})
\prod_{l=p+1}^{p+q}(D-\tilde{j}_{l})^{-1}\frac{\delta f}{\delta L^{N}}
\prod_{i=1}^{j}(D+j_{i}) \nonumber\\
& & \cdot (\prod_{i=j+1}^{p}(D+j_{i})
\prod_{l=p+1}^{p+q}(D-\tilde{j}_{l})^{-1}\frac{\delta g}{\delta L^{N}}
\prod_{i=1}^{j}(D+j_{i}))_{+} \nonumber\\
& & - \prod_{i=j}^{p}(D+j_{i})
\prod_{l=p+1}^{p+q}(D-\tilde{j}_{l})^{-1}\frac{\delta f}{\delta L^{N}}
\prod_{i=1}^{j-1}(D+j_{i}) \nonumber\\
& & \cdot (\prod_{i=j}^{p}(D+j_{i})
\prod_{l=p+1}^{p+q}(D-\tilde{j}_{l})^{-1}\frac{\delta g}{\delta L^{N}}
\prod_{i=1}^{j-1}(D+j_{i}))_{+}\} \nonumber\\
& & - \oint \sum_{k=p+1}^{p+q}Res\{\prod_{l=k}^{p+q}(D-\tilde{j}_{l})^{-1}
\frac{\delta f}{\delta L^{N}}\prod_{i=1}^{p}(D+j_{i})
\prod_{l=p+1}^{k-1}(D-\tilde{j}_{l})^{-1} \nonumber\\
& & \cdot (\prod_{l=k}^{p+q}(D-\tilde{j}_{l})^{-1}
\frac{\delta g}{\delta L^{N}}\prod_{i=1}^{p}(D+j_{i})
\prod_{l=p+1}^{k-1}(D-\tilde{j}_{l})^{-1})_{+} \nonumber\\
& & - \prod_{l=k+1}^{p+q}(D-\tilde{j}_{l})^{-1}
\frac{\delta f}{\delta L^{N}}\prod_{i=1}^{p}(D+j_{i})
\prod_{l=p+1}^{k}(D-\tilde{j}_{l})^{-1} \nonumber\\
& & \cdot (\prod_{l=k+1}^{p+q}(D-\tilde{j}_{l})^{-1}
\frac{\delta g}{\delta L^{N}}\prod_{i=1}^{p}(D+j_{i})
\prod_{l=p+1}^{k}(D-\tilde{j}_{l})^{-1})_{+}\}.
\end{eqnarray}
It is easy to see that only the first or last term in each part of (13)
survives the summations, and (13) turns out to be
\begin{eqnarray}
& & {\{\oint f(L^{N}), \oint g(L^{N})\}} \nonumber\\
&=& \oint Res[\frac{\delta f}{\delta L^{N}}\prod_{i=1}^{p}(D+j_{i})
\prod_{l=p+1}^{p+q}(D-\tilde{j}_{l})^{-1}(\frac{\delta g}{\delta L^{N}}
\prod_{i=1}^{p}(D+j_{i})\prod_{l=p+1}^{p+q}(D-\tilde{j}_{l})^{-1})_{+}
\nonumber\\
& & - \prod_{i=1}^{p}(D+j_{i})\prod_{l=p+1}^{p+q}(D-\tilde{j}_{l})^{-1})_{+}
\frac{\delta f}{\delta L^{N}}(\prod_{i=1}^{p}(D+j_{i})
\prod_{l=p+1}^{p+q}(D-\tilde{j}_{l})^{-1}\frac{\delta g}{\delta L^{N}})_{+}]
\nonumber\\
&=& \oint Res[\frac{\delta f}{\delta L^{N}}L^{N}
(\frac{\delta g}{\delta L^{N}}L^{N})_{+} - \frac{\delta f}{\delta L^{N}}
(L^{N}\frac{\delta g}{\delta L^{N}})_{+}L^{N}] \nonumber\\
&=& \oint Res[\frac{\delta f}{\delta L^{N}}K(\frac{\delta g}{\delta L^{N}})]
\end{eqnarray}
where the equality $\oint Res{[P,Q]}=0$ for arbitrary $P$, $Q$ has been
frequently used. Eq.(14) is none but (5) obtained in terms of (7), thus the
$\hat{W}_{1+\infty}^{(N)}$ realizations (8) hold.

The above derivation remains valid after an arbitrary mixing of the orders of
the $(D+j_{i})$ and $(D-\tilde{j}_{l})^{-1}$ terms in (8). It gives rise to
modified $p+q$ free boson representations of $\hat{W}_{1+\infty}^{(N)}$,
in response to different screening charge couplings of $j_{i}'$ and
$\tilde{j}_{l}'$ in the spin-2 stress tensor $v_{-N+1}$.

In terms of eq.(8), all $\hat{W}_{1+\infty}^{(N)}$ algebras are mutually
reducible. Namely, the $p-1$ plus $q$ boson representation of
$\hat{W}_{1+\infty}^{(N-1)}$ can be obtained from the $p$ plus $q$ boson
representation of $\hat{W}_{1+\infty}^{(N)}$, by simply, e.g., setting
$j_{1}=0$ and multiplying $L^{N}$ by $D^{-1}$ from the left:
\begin{eqnarray}
L^{(p-1)-q} = D^{-1}L^{p-q}\mid_{j_{1}=0} = \prod_{i=2}^{p}(D+j_{i})
\prod_{l=p+1}^{p+q}(D-\tilde{j}_{l})^{-1};
\end{eqnarray}
and in turn the latter may be obtained similarly from the $p$ plus $q+1$
boson representation of $\hat{W}_{1+\infty}^{(N-1)}$:
\begin{eqnarray}
L^{p-q} = L^{p-(q+1)}D\mid_{\tilde{j}_{p+q+1}=0} = \prod_{i=1}^{p}(D+j_{i})
\prod_{l=p+1}^{p+q}(D-\tilde{j}_{l})^{-1}.
\end{eqnarray}
The point is that, because $j_{1}$ and $\tilde{j}_{p+q+1}$ are {\it free}
currents, neither constraint effects the Poisson brackets (7) of the rest
of the currents so that the composite $W$-structures from (15)-(16) can
be naturally derived and coincides with (5).

The realizations of $\hat{W}_{\infty}^{(N)}$ are acquired by just imposing
the second-class constraint
\begin{eqnarray}
v_{-N}=\sum_{i}j_{i}+\sum_{l}\tilde{j}_{l}=0
\end{eqnarray}
on (8) (and on the versions with $j_{i}$ and $\tilde{j}_{l}$ terms mixed).
They appear, without loss of elegency, as
\begin{eqnarray}
L^{N} = \prod_{i=1}^{p}(D+\vec{h}_{i}\cdot\vec{j})
\prod_{l=p+1}^{p+q}(D-\vec{h}_{l}\cdot\vec{j})^{-1}
\end{eqnarray}
where we denote the bosonic currents as a vector $\vec{j}=(j_{1},\cdots,
j_{p},-\tilde{j}_{p+1},\cdots,-\tilde{j}_{p+q})$ projected onto the
overcomplete set $(\vec{h}_{i},\vec{h}_{l})$, that have the inner product
\begin{eqnarray}
\vec{h}_{i}\cdot\vec{h}_{j} = \delta_{ij}-\frac{1}{p-q},
{}~~~ \vec{h}_{l}\cdot\vec{h}_{k} = -\delta_{lk}-\frac{1}{p-q},
{}~~~ \vec{h}_{i}\cdot\vec{h}_{l} = \frac{1}{p-q},
\end{eqnarray}
obeying $\sum_{i}\vec{h}_{i}+\sum_{l}\vec{h}_{l}=0$. An equivalent
statement is that the Poisson brackets (7) are modified to
\begin{eqnarray}
& & {\{j_{i}(z),j_{j}(z')\}} = (\delta_{ij}-\frac{1}{p-q})
\partial_{z}\delta(z-z'), \nonumber\\
& & {\{\tilde{j}_{l}(z),\tilde{j}_{k}(z')\}} = -(\delta_{lk}+\frac{1}{p-q})
\partial_{z}\delta(z-z'), \nonumber\\
& & {\{j_{i}(z),\tilde{j}_{l}(z')\}} = \frac{1}{p-q}\partial_{z}\delta(z-z').
\end{eqnarray}

It is obvious that eq.(18) or (8) are a set of generalizations of the Miura
transformation (1) for $W_N$ or
\begin{eqnarray}
Q = D^{N} + \sum_{r=0}^{N-1}q_{r}D^{r} = \prod_{i=1}^{N}(D+j_{i})
\end{eqnarray}
for $W_{N+1}$. With $\tilde{j}_{l}=j_{i\geq p-q+1}=0$, the former reduces
to the latter, and the Dirac brackets for the rest of the currents appear
exactly the same as the Poisson brackets. From the Hamiltonian point of view,
the $W_{N+1}$ ($W_{N}$) embeddings to (6) are natural, and the constraints
$v_{r\geq 0}=0$, which are at least linear in $\tilde{j}_{l}$ or
$j_{i\geq p-q+1}$, are actually first class. Combining with the results of
last two paragraphs, $\hat{W}_{1+\infty}^{(N)}$ ($\hat{W}_{\infty}^{(N)}$)
all serve as universal nonlinear $W$-algebras.

The many boson realizations (18) are also generalizations of the two boson
realization (2) of $\hat{W}_{\infty}$, though the latter looks a little
different. In particular, with $p=2,q=1$, the mixed version of (18)
\begin{eqnarray}
L &=& (D+\vec{h}_{1}\cdot\vec{j})(D-\vec{h}_{3}\cdot\vec{j})^{-1}
(D+\vec{h}_{2}\cdot\vec{j}) \nonumber\\
&=& (D+j_{1})\frac{1}{D+j_{1}+j_{2}}(D+j_{2})~=~
D+j_{2}\frac{1}{D+j_{1}+j_{2}}j_{1}
\end{eqnarray}
is identical to (2) by setting $j_{2}\equiv -\bar{j}$, $j_{1}\equiv -j$.
Note the $\bar{j}$, $j$ Poisson brackets from (20) are precisely those for
a pair of complex free bosons.

Eq.(8) of (18), etc., are automatically various many boson realizations of
the KP operator, for example,
\begin{eqnarray}
L = (D+j_{1})\frac{1}{(D-\tilde{j}_{2})}(D+j_{3})\frac{1}{(D-\tilde{j}_{4})}
\cdots \frac{1}{(D-\tilde{j}_{2s})}(D+j_{2s+1}).
\end{eqnarray}
Since $\hat{W}_{1+\infty}^{(N)}$ ($\hat{W}_{\infty}^{(N)}$) are just KP
Hamiltonian structures, the KP hierarchy, which is generated by the involutive
Hamiltonian charges $\oint H_{m}(z)=\frac{1}{m}\oint Res L^{m}$ $(m\geq 1)$,
\begin{eqnarray}
\frac{\partial L}{\partial t_{m}} = {[(L^{m})_{+}, L]}
\end{eqnarray}
itself enjoys these realizations. The number of bosons (of different
signatures) can be arbitrarily large, thus the
realizations are faithful at least formally. We call them the modified KP
hierarchies, containing all the modified KdV hierarchies [5] by reduction.
On the other hand, we may directly propose a set of well-defined
integrable $\vec{j}\equiv (j_{i},\tilde{j}_{l})$-hierarchies
\begin{eqnarray}
\frac{\partial \vec{j}}{\partial t_{m}} = {\{\vec{j}, \oint H_{m}\}}
\end{eqnarray}
with the right side evaluated according to the Poisson brackets of $\vec{j}$.
By composition in terms of the realizations, (25) leads to (the equivalent
Hamiltonian form of) the modefied KP hierarchies. It enables us to
have explicit field-theoretical studies of these hierarchies.

Let us in turn seek a Lie-algebraic interpretation of the
$\hat{W}_{\infty}^{(N)}$ algebras along with that of the KP hierarchy. In
fact, this has been implied in eqs.(18-19). We now verify that the set
$(\vec{h}_{i},\vec{h}_{l})$ satisfying (19) are the weight vectors of
the fundamental representation of the graded Lie algebra $SL(p,q)$. Let
$\epsilon_{i}$ $(i=1,\ldots,p)$ and $\delta_{l}$ $(l=p+1,\ldots,p+q)$ are
orthonormal bases of $p+q$ dimensional linear space with positive and
negative metrics respectively. The simple roots of $SL(p,q)$, $\alpha_{s}$
$(s=1,\ldots,p+q-1)$, are chosen as (upto Weyl-group symmetry)
\begin{eqnarray}
& & \alpha_{2m-1}=\epsilon_{m}-\delta_{m},~~
\alpha_{2m}=\delta_{m}-\epsilon_{m+1},~~~(m=1,\ldots,q), \nonumber\\
& & \alpha_{2q+n}=\epsilon_{q+n}-\epsilon_{q+n+1}, ~~~(n=1,\ldots,p-q-1).
\end{eqnarray}
The fundamental weights $\lambda_{s}$ $(s=1,\ldots,p+q-1)$, satifying
$\lambda_{s}\cdot\alpha_{t}=\delta_{st}$, are correspondingly
\begin{eqnarray}
& & \lambda_{2m-1}=\sum_{i=1}^{q}(1-\frac{1}{p-q})\alpha_{2i-1}+
\sum_{i=m}^{q}\alpha_{2i}+\sum_{i=1}^{p-q-1}(1-\frac{i}{p-q})\alpha_{2q+i},
\nonumber\\
& & \lambda_{2m}=\sum_{i=1}^{m}\alpha_{2i-1}, \\
& & \lambda_{2q+n}=\sum_{i=1}^{q}(1-\frac{n}{p-q})\alpha_{2i-1}+
\sum_{i=1}^{n}i(1-\frac{n}{p-q})\alpha_{2q+i}+\sum_{i=n+1}^{p-q-1}
n(1-\frac{i}{p-q})\alpha_{2q+i}. \nonumber
\end{eqnarray}
It follows that the $p+q$ weight vectors $h_{s}=\lambda_{s}-\lambda_{s-1}$
$(\lambda_{0}=\lambda_{p+q}=0)$ are identical to $(\vec{h}_{i},\vec{h}_{l})$
in (19). Therefore characteristically, $\hat{W}_{\infty}^{(N)}$ are
associated with the graded $SL(p,q)$ with $p-q=N$ in terms of the KP Lax
operator (18).

An immediate implication is that one may, {\it \`a la} Drinfeld-Sokolov [1],
reconstruct the (modified) KP hierarchy as well as its Hamiltonian
structures $\hat{W}_{\infty}^{(N)}$ according to the (level-1)
$SL(p,q)\equiv A(p-1,q-1)$ Kac-Moody algebras. One may also obtain different
sets of integrable hierarchies of the KP type and $\hat{W}$-infinity algebras
from other series of graded Kac-Moody algebras. Meanwhile, the $SL(p,q)$
origin of $\hat{W}_{\infty}^{(N)}$ allows us to
carry out the quantization of the latter systematically through the
Drinfeld-Sokolov reduction or the Casimir construction [16], besides from the
transformation (18) directly. Moreover, with the prescription suggested in
Ref.[17], one expects to establish solvable nonlinear
$\hat{W}_{\infty}^{(N)}$-gravity models via quantum Hamiltonian reduction of
gauged $SL(p,q)$ WZW models. Finally, we note that by associating the
$SL(p,p-1)$ fundamental weight vectors to the Lax operator of a set of
scalar superfields, instead of scalars of different signatures, one gets
super $W$-algebras of the finite type [18]. Thus naturally, {\it heterotic}
$W$-algebras would arise via relevant $SL(p,q)$ Lax operators. We shall
address these problems elsewhere.

\vspace{10 pt}
\begin{center}
{\large Acknowledgements}
\end{center}

The author thanks P. van Nieuwenhuizen and Y.-S. Wu for discussions, and
is grateful to C. N. Yang for providing a stimulating atmosphere and support.
This work is supported in part by NSF grant PHY-9008452.

\vspace{10 pt}
\begin{center}
{\large References}
\end{center}
\begin{itemize}

\item[1.] V. Drinfeld and V. Sokolov, Journ. Sov. Math. 30 (1985) 1975.
\item[2.] A. B. Zamolodchikov, Theor. Math. Phys. 65 (1985) 1205.
\item[3.] F. Yu and Y.-S. Wu, Phys. Rev. Lett. 68 (1992) 2996.
\item[4.] A. B. Zamolodchikov and V. A. Fateev, Nucl. Phys. B280 [FS18]
(1987) 644; V. A. Fateev and S. L. Lykyanov, Int. J. Mod. Phys. A3 (1988) 507.
\item[5.] B. A. Kupershmidt and G. Wilson, Invent. Math. 62 (1981) 403.
\item[6.] F. Yu and Y.-S. Wu, Nucl. Phys. B373 (1992) 713.
\item[7.] I. Bakas, Phys. Lett. B228 (1989) 57; C. Pope, L. Romans and
X. Shen, Phys. Lett. B236 (1990) 173.
\item[8.] L. A. Dickey, Annals New York Academy of Sciences, 491 (1987) 131.
\item[9.] M. Sato, RIMS Kokyuroku 439 (1981) 30; E. Date, M. Jimbo, M.
Kashiwara and T. Miwa, in Proc. of RIMS Symposium on Nonlinear Integrable
Systems, eds. M. Jimbo and T. Miwa, (World Scientific, Singapore, 1983);
G. Segal and G. Wilson, Publ. IHES 61 (1985) 1.
\item[10.] I. Bakas and E. Kiritsis, Maryland/Berkeley/LBL preprint
UCB-PTH-91/44, LBL-31213 or UMD-PP-92/37, Sept. 1991.
\item[11.] F. Yu and Y.-S. Wu, Phys. Lett. B294 (1992) 177.
\item[12.] F. Yu and Y.-S. Wu, Utah preprint UU-HEP-92/12; 92/13.
\item[13.] J. M. Figueroa-O'Farrill, J. Mas and E. Ramos, preprint
BONN-HE-92/20; 92/$\alpha$.
\item[14.] V. G. Kac, Adv. Math. 26 (1977) 8.
\item[15.] F. Yu, Stony Brook/Utah preprint ITP-SB-92-65/UU-HEP-92/22.
\item[16.] F. A. Bais, P. Bouwknegt, M. Surridge and K. Schoutens, Nucl.
Phys. B304 (1988) 348; 371.
\item[17.] K. Schoutens, A. Sevrin and P. van Nieuwenhuizen, Nucl. Phys.
B364 (1991) 584; B371 (1992) 315.
\item[18.] K. Ito, Nucl. Phys. B370 (1992) 123.

\end{itemize}

\end{document}